\documentclass[10pt, final]{IEEEtran}

\usepackage{amsthm}
\usepackage{amsmath}
\usepackage{bbm}
\usepackage[ruled]{algorithm2e}
\usepackage{mathrsfs}
\usepackage{cite}
\usepackage{bm,epsfig,amsthm,url}
\usepackage{indentfirst}
\usepackage{amssymb}
\usepackage{amsfonts}
\usepackage{epstopdf}
\usepackage[caption=false,font=footnotesize]{subfig}
\usepackage{xcolor}
\usepackage{mathtools}

\usepackage{hyperref}
\usepackage{balance} 

\theoremstyle{definition}

\newtheorem{proposition}{Proposition}

\newtheorem{remark}{Remark}

\usepackage{graphicx}
\usepackage{epstopdf}

\begin{document}
%\pagewiselinenumbers

\title{Intelligent Reflecting Surface Assisted Multi-Cluster AirComp via Dynamic Beamforming}

\author{
%	Yapeng~Zhao, 
%	Qingqing~Wu,~\IEEEmembership{Senior Member,~IEEE},
%	Wen~Chen,~\IEEEmembership{Senior Member,~IEEE},
%	Celimuge~Wu,~\IEEEmembership{Senior Member,~IEEE,}, 
%	Julian~Cheng,~\IEEEmembership{Fellow,~IEEE},
%	Octavia~A.~Dobre,~\IEEEmembership{Fellow,~IEEE},
	Yapeng~Zhao,~Qingqing~Wu,~Wen~Chen,~Celimuge~Wu,~and~Octavia~A.~Dobre,~\IEEEmembership{Fellow,~IEEE}\hspace{-5pt}
\thanks{Y. Zhao is with the State Key Laboratory of Internet of Things for Smart City, University of Macau, Macao 999078, China, and also with the Institute for Signal Processing and Systems at Shanghai Jiao Tong University, 200240, China (email: yc17435@connect.um.edu.mo).  
Q. Wu and W. Chen is with the Department of Electronic Engineering, Shanghai Jiao Tong University, 200240, China (e-mail: qingqingwu@sjtu.edu.cn; wenchen@sjtu.edu.cn). 
C. Wu is with Graduate School of Informatics and Engineering, The University of Electro-Communications, Tokyo,182-8585, Japan (e-mail: Celimuge@uec.ac.jp).
O. A. Dobre is with the Faculty of Engineering and Applied Science, Memorial University, St. John’s, NL A1B 3X5, Canada (e-mail:
odobre@mun.ca).
}
}

\maketitle

\begin{abstract}

This paper studies an multi-cluster over-the-air computation (AirComp) system, where an intelligent reflecting surface (IRS) assists the signal transmission from devices to an access point (AP). 
The clusters are activated to compute heterogeneous functions in a time-division manner.
Specifically, two types of IRS beamforming (BF) schemes are proposed to reveal the performance-cost tradeoff. One is the cluster-adaptive BF scheme, where each BF pattern is dedicated to one cluster, and the other is the dynamic BF scheme, which is applied to any number of IRS BF patterns. By deeply exploiting their inherent properties, both generic and low-complexity algorithms are proposed in which the IRS BF patterns, time and power resource allocation are jointly optimized. Numerical results show that IRS can significantly enhance the function computation performance, and demonstrate that the dynamic IRS BF scheme with half of the total IRS BF patterns can achieve near-optimal performance which can be deemed as a cost-efficient approach for IRS-aided multi-cluster AirComp systems.
\end{abstract}
%\vspace{-6mm}

\section{Introduction}

The sixth generation (6G) network aims to integrate communication, sensing, computing, and intelligence together, thus constructing native multi-functional systems that serve as powerful engines to build an intelligent world \cite{Towards6G, YuanmingShi_JSAC22_EdgeAI, XiaoyangLi_Integrated_SensingComputation}. A series of advanced services are progressively conceived and experimented in such multi-functional systems, such as extended reality (XR), edge artificial intelligence (AI), and autonomous driving \cite{WalidSaad_6GVision, YuanmingShi_Survey20_EdgeAI}, either of which needs to collect and process enormous data from distributed devices. Conventionally, the raw data collection and process are regarded as separate procedures and designed in isolation. 

%The traditional multiple-access (MA) strategies for data collection from tremendous devices are deemed task-agnostic \cite{AirComp_survey}, which aim to realize reliable data transmission as fast as possible. 

In practice, these intelligent services demand the functional computation results of the raw data, such as the maximum reading of several temperature monitors, the average of local models, and the minimum distance between cars and barricades. The recently aroused over-the-air computation (AirComp), deemed as a task-oriented multiple-access (MA) strategy, has shown its high efficiency in integrating communication and computation through leveraging the signal superposition characteristic of the wireless channel \cite{AirComp_survey}. The co-channel interference is treated as a catalyst in AirComp for computation, whereas traditional task-agnostic MA techniques consider it to be detrimental.
%The computable function space in AirComp was extended from simple linear functions to arbitrary concrete functions through nomographic decomposition \cite{AirComp_survey}, and further incorporated any implicit functions that can be approximated via deep neural networks in a data-driven manner \cite{DeepAirComp}. 
AirComp was conceived \cite{KaibinHuang_IoT19_MIMO_AirComp, YapengZhao} and validated \cite{AirComp_Implement21} to be a promising scalable function computation strategy in a series of works. Specifically, the transceiver design that exhibits a uniform-forcing structure to compensate individual channel fading was proposed in \cite{LiChen_UniformForcing}, and the multiple-input multiple-output (MIMO) AirComp \cite{KaibinHuang_IoT19_MIMO_AirComp} was exploited for simultaneously multi-modal sensing. 
%The authors in \cite{AirComp_Implement21} analyzed the effect of frame timing offset and carrier frequency offset, and further implemented the prototype of AirComp based federated learning (FL) system. 
However, the harsh wireless propagation environment still significantly weakens the signal strength and thus deteriorates the computation performance.

Intelligent reflecting surface (IRS), a promising technology for future wireless networks, has shown the potential to overcome this detrimental effect by adapting the phase shifts of low-cost passive elements. Several works have exhibited the effectiveness of IRSs in the single-cluster AirComp system, e.g., the authors in \cite{YuanmingShi_TWC21_FL_IRS, TwoTimescaleIRS_AirComp} employed an IRS to strengthen channels thus reduce the distortion. If the single-cluster AirComp is intuitively extended to a multi-cluster case like the conventional communication systems \cite{Mario_TWC15_Clustered, AirComp_survey}, the considered configuration times of IRS passive beamforming (BF) is the same as the number of participated clusters, otherwise its BF remains static during a transmission frame, i.e., each cluster is assigned with a dedicated BF pattern or all clusters employ the same pattern. It is observed that the two policies are rigid in that they provide the upper and lower bounds on system performance with no tradeoff. 
Although harnessing more BF patterns provides more degrees of freedom (DoFs) which can configure a suitable propagation environment for devices, thereby further enhancing the system performance, it also adds extra signaling overhead. 
Hence, a flexible dynamic IRS BF scheme is desired to balance the system  performance and cost. 
Recently, a novel dynamic IRS beamforming (DIBF) technique has been proposed to offer promising DoF thus enabling more flexible resource allocation in half-duplex and full-duplex wireless powered communication networks (WPCNs) \cite{DynamicWPCNs, MengHuaDynamic}, mobile edge computing (MEC) with binary offloading \cite{GuangjiChenDynamic}, etc. 
DIBF proactively develops favorable time-selective channels by harnessing the unique character of IRSs that the BF patterns can be tuned multiple times within a given channel coherence time. Additionally, a good balance between performance, associated tuning costs, and signaling overhead can be achieved by flexibly managing the amount of reconfigurations.

In this work, we analyze an IRS-aided AirComp system comprising of multiple clusters by taking the performance-cost tradeoff into account. 
The previous studies concentrated on the transceiver design in one dedicated time slot (TS) that aim to minimize the function computation error measured by the mean square error (MSE).
We study in this paper how to boost the computation capability within a given period in which the IRS can be configured multiple times. 
Specifically, clusters are activated in a time-division manner to eliminate the inter-cluster interference thereby computing heterogeneous functions, and we present two types of BF schemes to unleash the fundamental performance-cost tradeoff.
The cluster-adaptive BF method, where each BF is allocated to one cluster, is first taken into consideration to provide the upper bound for comparison. The performance-cost tradeoff is then explored using the general dynamic IRS BF method, which is be applied to any number of BF patterns. 
%We propose two efficient algorithms based on difference-of-convex (DC) and successive convex approximation (SCA) techniques to solve two problems. To reduce the computational complexity, we further propose an efficient algorithm by deeply exploiting the inherent property of dynamic BF configuration scheme. 
Simulation results verify the theoretical findings and show how IRS improves the performance of multi-cluster AirComp. In particular, it is observed that the proposed DIBF design is sufficient to acquire near-optimal performance with a minimal computation rate decrease (less than $6\%$) by halving the number of BF patterns, which sheds light on balancing the performance-cost via limiting the number of BF patterns.

\vspace{-5pt}
\begin{figure}[t]
	\centering
	\includegraphics[width=0.65\columnwidth]{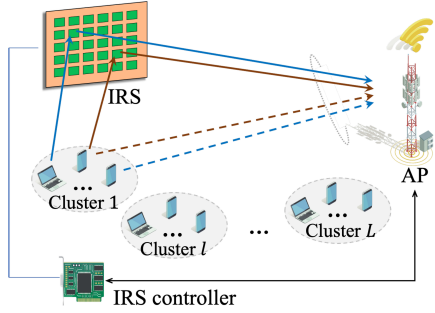}
	\caption{An IRS-assisted multi-cluster AirComp system.}
	\label{fig_AirComp_DIBF}
\end{figure}

%\vspace{-5mm}
\begin{figure}[t]
	\centering
\includegraphics[width=0.95\columnwidth]{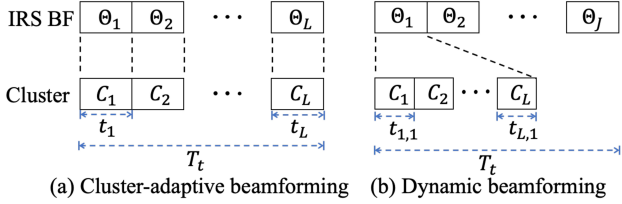}
	\caption{Transmission protocol of the multi-cluster AirComp with DIBF.}
	\label{fig_DIBF}
\end{figure}

%%%%%%%%%%%%%%%%%%%%%%%%%%%%%%%%%%%%%%%%%%%%%%%%
\section{System Model and Problem Formulation}
As depicted in Fig. \ref{fig_AirComp_DIBF}, the considered multi-cluster AirComp system composes of a single antenna access point (AP), an IRS equipped with $N$ passive elements, and $K$ devices. Due to the different computational requirements, the $K$ devices are divided into $L$ clusters. Each cluster consists of $K_{\ell}$ devices and computes a desired  type-$\ell$ function named $f_{\ell}$. We denote $\mathcal{K}_\ell\triangleq \{1, \ldots, K_\ell\}$ as the set of devices within $\ell$-th cluster, where $\mathcal{K}_{i} \cap \mathcal{K}_{j}= \varnothing, \forall i \neq j,\; i, j \in \mathcal{L},\; \mathcal{L} = \{1,\ldots,L\}$. The AP computes heterogeneous functions via AirComp from each cluster in a time-division manner.
Specifically, the AP computes $f_\ell$ with the raw data from all the $K_{\ell}$ devices in cluster $\ell$. Let $s_{k,\ell}$ denote one dedicated reading of the device $k$ in cluster $\ell$. The target type-$\ell$ function computed at AP is expressed as
\begin{align}
	f_\ell =  \phi_\ell \left(\sum\nolimits_{k\in \mathcal{K}_\ell} \varphi_k(s_{k,\ell}) \right),
\end{align} 
where $\varphi_k(\cdot)$ is the pre-processing function at device $k$, and $\phi_\ell(\cdot)$ is the post-processing function at the AP. We adopt the digital AirComp to perform function computation, in which the nest lattice coding strategy is harnessed to achieve the integer combinations of transmitted codewords and combat the channel noise meanwhile. Specifically, the reading of each device is first processed by $\varphi_k(\cdot)$, then the result is quantized and mapped into a vector, and finally, it is encoded into a nested lattice codeword (see \cite{Nazer_TIT11, Mario_TWC15_Clustered} for more details). We assume that each symbol $x_{k}$ of the transmitted signal has been normalized into a mutual independent symbol with zero mean and unit power. Hence, the target recovered data at the AP of cluster $\ell$ when computing a type-$\ell$ function is given by $y_\ell =  \sum_{k\in \mathcal{K}_\ell} x_{k,\ell}$.

The baseband channels from the IRS to AP, from the device $k$ to IRS, and from the device $k$ to AP are denoted by $\bm{g}^{\mathrm{H}} \in \mathbb{C}^{1 \times N}$, $\bm{h}^r_{\ell,k} \in \mathbb{C}^{N \times 1}$ and ${h}^d_{\ell,k} \in \mathbb{C}$, respectively. 
In this paper, we assume that all the involved channels are perfectly known by applying the state-of-the-art channel estimation techniques  
\cite{WQQ_IRS_Survey}.
All devices in each cluster transmit concurrently with the symbol-level synchronization assumed. In a channel coherence interval $T_t$, the duration required to compute a function is further divided into $J$ TSs. In addition, $t_j$, where $\mathbf{\Theta}_j$ is employed in, denotes the time interval for the $j$-th TS. 
Then, $t_j$ is divided to $L$ sub-TSs for the function computation in different clusters.
In sub-TS $t_{\ell,j}$, the recovered signal from cluster $\ell$ to compute $f_{\ell}$ is given by
\begin{align}
	\hat{y}_{\ell,j} &= a_{\ell,j} \bigg( \sum_{k\in \mathcal{K}_\ell} {h}^e_{\ell,j,k} b_{\ell,j,k} {x}_{\ell,j,k} + {z}_{\ell} \bigg)  \nonumber \\
	&= \!\! \sum_{k\in \mathcal{K}_\ell} \! {x}_{\ell,j,k} \! + \!\!\! \sum_{k\in \mathcal{K}_\ell} \!\! \big(a_{\ell,j} {h}^e_{\ell,j,k} b_{\ell,j,k} \! - \! 1\big){x}_{\ell,j,k} \! + \!a_{\ell,j}z_{\ell},
\end{align} 
where $b_{\ell,j,k}, a_{\ell,j} \in \mathbb{C} $ are the transmitter and the receiver scalars, respectively. Besides, ${h}^e_{\ell,j,k} = {h}^d_{\ell,k} + \bm{g}^{\mathrm{H}}_{\ell} \bm{\Theta}_{j} \bm{h}^r_{\ell,k} = {h}^d_{\ell,k} + \bm{v}_{j}^{\mathrm{H}} \mathrm{diag}(\bm{g}_{\ell}^{\mathrm{H}}) \mathbf{h}^r_{\ell,k}$ denotes the composite device-AP channel coefficient of device $k$ in cluster $\ell$, where $\bm{\Theta}_j = \operatorname{diag}(e^{\jmath\theta_{j,1}}, \ldots, e^{\jmath\theta_{j,N}} ) $ denotes the phase shift matrix of IRS, in which $\theta_{j,n} \in [0,2\pi), \forall n \in \mathcal{N}$ denotes the phase shift on the incident signal of $n$-th element \cite{WQQ_TWC19_IRS}, $\bm{v}_{j} = [v_{j,1}, \ldots, v_{j,N}]^{\mathrm{H}}$ with $v_{j,n} = e^{\jmath \theta_n}$, and $z_{\ell} \in \mathbb{C} \sim \mathcal{CN}(0, \sigma_{\ell}^2) $ denotes the additional white Gaussian noise.
%$\bm{n} \in \mathbb{C}^M \sim \mathcal{CN}(0, \sigma^2 \bm{I}) $ denotes the additional white Gaussian noise. 

%Besides, we assume that all the devices are energy-constrained, i.e., $\sum_j t_{\ell,j}|b_{\ell,j,k}|^2\leq E_{\max}, \forall k$.

The landmark work in \cite{Nazer_TIT11} established the fundamentals of \textit{compute-and-forward} strategy from the information theory perspective. 
Accordingly, the subsequent work \cite{Mario_TWC15_Clustered} proposed the computation rate of AirComp, defined as the achieved number of computed function values per channel use (in num/Hz), which can be expressed as
\begin{align} \label{computation_rate} 
	r_{\ell,j}^{\mathrm{c}}=\frac{1}{m_{\ell}+\log _{2} K_{\ell}} \log _{2}^+\left(\frac{1}{\mathrm{MSE}(\hat{y}_{\ell,j}, y_{\ell,j})}\right),
\end{align}
with $m_\ell$ being the number of the optimal quantization bits for the type-$\ell$ function computation, $K_\ell$ being the number of involved devices, and $\log _{2}^+(\cdot) \triangleq \max\{\log _{2}(\cdot),0\}$. It can be observed from \eqref{computation_rate} that the computation rate maximization problem can be equivalently transformed into the MSE minimization problem. 
By adopting the uniform-forcing transceiver proposed in \cite{LiChen_UniformForcing}, the solution to the MSE minimization problem is
\begin{align} \label{e_MSE_opt1}
	\! a_{\ell,j} \!\! = \!\! \frac{1}{\sqrt{\eta_{\ell,j}}}, b_{\ell,j,k} \!\! = \!\! \sqrt{\eta_{\ell,j}}\frac{(h_{\ell,j,k}^e)^{\mathrm{H}}}{|h_{\ell,j,k}^e|^2}, \eta_{\ell,j} \!\! = \!\!  P_0 \min_k |h_{\ell,j,k}^e|^2, \!
\end{align}
where $P_0$ denotes the maximum allowed transmit power. Similarly, for the case when devices are energy-limited, we have
\begin{align} \label{e_MSE_opt2}
	\eta_{\ell,j} = \min_k (p_{\ell,j,k}|h_{\ell,j,k}^e|^2),
\end{align}
where $ p_{\ell,j,k}$ is the allocated power of device $k$ for current transmission. The corresponding computation rate of cluster $\ell$ under BF pattern $\mathbf{\Theta}_j$ is given by
\vspace{-2pt}
\begin{align}
	r_{\ell,j}^{\mathrm{c}}=\frac{1}{m_{\ell}+\log _{2} K_{\ell}} \log _{2}^+\left(\frac{ \min_k (p_{\ell,j,k}|h_{\ell,j,k}^e|^2)}{ \sigma^2_\ell }\right).
\end{align}
%Accordingly, the achievable computation rate of cluster $\ell$ in a channel coherence interval is given by $ R_{\ell}^{\mathrm{c}} = \sum_{j=1}^J r_{\ell,j}^{\mathrm{c}} $.
%$$ \sum\nolimits_{j=1}^J, \sum\limits_{j=1}^J$$

%\begin{align}
%	\bm{g} ^{\mathrm{H}}\bm{\Theta}_{\ell} \bm{h}^r_{\ell,k} =  \bm{v}_{\ell}^{\mathrm{H}} \mathrm{diag}(\bm{g}^{\mathrm{H}}) \bm{h}^r_{\ell,k}
%\end{align}
%
%$\bm{v}_{\ell}^{\mathrm{H}} = \{v_{\ell,1}, \ldots,  v_{\ell,N}\}$

We aim to boost the computation capability within $T_t$ in which IRS can be configured multiple times.  As shown in Fig.~\ref{fig_DIBF}, the two proposed IRS BF schemes depend on how the IRS sets its BF during the time interval, and are specified as follows.
%One is the cluster adaptive IRS BF, i.e., the IRS is allowed to reconfigure its BF patterns $L$ times and each BF is dedicated to one cluster, the other is dynamic IRS BF shown in Fig. \ref{fig_DIBF}, i.e., the IRS is allowed to reconfigure its BF patterns $J$ times and can be flexibly associated to clusters. The two types of IRS BF configuration are specified as follows.

%%%%%%%%%%%%%%%%%%%%%%%%%%%%%%%%%%%%%%%%%%%%%%%%
\subsection{Cluster-adaptive IRS BF}
Each cluster is mapped with a unique IRS BF pattern as well as the TSs in the cluster-adaptive BF scheme, which provide the performance upper bound, i.e., arbitrary BF schemes with $J > K$ cannot further improve the performance~\cite{DynamicWPCNs}. 
Accordingly, the computation rate maximization problem is given by 
\vspace{-2pt}
\begin{subequations}
	\begin{align}
		\!\!\!\!\!\text{(P1)}: \!\!\mathop  {\max }\limits_{\scriptstyle \{t_{\ell}\}, \{\bm{v}_{\ell}\}, \atop \scriptstyle 
			\{p_{\ell,k}\}}  \;\; \!\! \!&\sum\limits_{\ell=1}^L  \!\! \frac{w_\ell t_{\ell}}{m_{\ell} \!+ \! \log _{2} K_{\ell}} \! \log _{2}^+\! \left( \!\! \frac{ \min_k (p_{\ell,k}|h_{\ell,k}^e|^2)}{ \sigma^2_\ell }\!\! \right)  \label{e_obj_P1} \\
		{\rm{s.t.}}\;\;\;\;\;\;
		& t_{\ell} \geq 0, \ p_{\ell,k} \geq 0, \ \forall \ell,k, \label{e_c_P1_t1} \\
		& \sum\nolimits_{\ell=1}^L t_{\ell} \leq T_t,   \label{e_c_P1_t2} \\
		& t_{\ell} p_{\ell,k} \leq E_{\max}, \ \forall \ell, k, \label{e_c_P1_tp} \\
		& |v_{\ell,n}| = 1,\ \forall \ell,n, \label{e_c_P1_v}
	\end{align}
\end{subequations}
where $w_\ell$ denotes the weight of cluster $\ell$. Since the weight $w_\ell$, quantization bit $m_\ell$, and the number of devices $K_\ell$ of each cluster do not affect the scheme design, we set that $w_\ell = 1, m_\ell = \tilde{m}, K_\ell = \tilde{K}, \forall \ell $. Note that (P1) is generally a non-convex optimization problem with non-convex norm-one constraint \eqref{e_c_P1_v} and the coupled variables in \eqref{e_obj_P1} and \eqref{e_c_P1_tp}.

\subsection{Dynamic IRS BF}

For the proposed general framework that the BF at IRS can be reconfigured $J$ times during $T_t$, the clusters can flexibly choose one or multiple BF patterns for computation. The corresponding problem is formulated as
\vspace{-2pt}
\begin{subequations}
\begin{align}
\!\!\!\text{(P2)}: \mathop  {\max }\limits_{\scriptstyle \{t_{\ell,j}\}, \{\bm{v}_{j}\}, \atop \scriptstyle 
\{p_{\ell,j,k}\}}  \;\; \!\! \!&\sum\nolimits_{\ell=1}^L \sum\nolimits_{j=1}^J w_\ell t_{\ell, j} r_{\ell, j}^{\mathrm{c}} \label{e_p2} \\ 
{\rm{s.t.}}\;\;\;\;\;\;
& t_{\ell,j} \geq 0,\  p_{\ell,j,k} \geq 0,\  \forall \ell, j,k , \label{e_c_P2_t1} \\
& \sum\nolimits_{\ell=1}^L \sum\nolimits_{j=1}^J t_{\ell, j} \leq T_t , \label{e_c_P2_t2} \\
& \sum\nolimits_{j=1}^J t_{\ell,j} p_{\ell,j,k} \leq E_{\max},\ \forall \ell, k, \label{e_c_P2_tp} \\
& |v_{j,n}| = 1,\ \forall j,n. \label{e_c_P2_v}
\end{align}
\end{subequations}
Similar to (P1), we assume that $w_\ell = 1, m_\ell = \tilde{m}, K_\ell = \tilde{K}, \forall \ell $. Note that (P2) is still non-convex since there exist highly-coupled variables in both \eqref{e_p2} and \eqref{e_c_P2_tp}, and constraint \eqref{e_c_P2_v} is generally non-convex.

\section{Proposed Algorithms}

\subsection{Proposed Algorithm for (P1)}

(P1) is shown as a maximin problem, which can be converted to a traditional maximization problem by intuitively introducing the auxiliary variables to substitute the inner minimization part. Then, multiple techniques can be adopted, such as alternating optimization. However, it is still inefficient and may suffer from performance loss. In this section, by analyzing the inherent properties of (P1), we propose an efficient algorithm correspondingly.

\begin{proposition} \label{proposition1}
The optimal solution of (P1) satisfies, 	
\begin{align}
	\min\nolimits_k (p_{\ell,k}^* |h^e_{\ell,k}|^2) = E_{\max}/t_\ell \min\nolimits_k |h^e_{\ell,k}|^2,\ \forall \ell.
	\end{align}
\end{proposition}

\quad {\it{Proof:}} Suppose that $\mathcal{S}^*= \big\{ \{t_{\ell}^*\}, \{\bm{v}_{\ell}^*\}, \{p_{\ell,k}^*\}\big\}$ achieves the optimal solution to (P1), we have $p_{\ell,k}^* \leq E_{\max}/t_\ell^*,\ \forall \ell,k$. The maximum allowed transmit power for the devices in cluster $\ell$ becomes $P_0 = E_{\max}/t_\ell^*,\ \forall \ell$. By combing it with \eqref{e_MSE_opt1}, Proposition~\ref{proposition1} is obtained. $\hfill\blacksquare$

Inspired by Proposition \ref{proposition1}, we introduce the auxiliary variables $\{\Gamma_{\ell}\}$ that satisfy $  \min_k |h^e_{\ell,k}|^2 \geq \Gamma_{\ell},\ \forall \ell$ and reformulate (P1)~as
\vspace{-5pt}
\begin{subequations} \label{P11}
\begin{align}
\mathop  {\max }\limits_{\scriptstyle \{t_{\ell}\}, \{\bm{v}_{\ell}\}, \atop \scriptstyle 
\{\Gamma_{\ell}\}}  \;\; \!\! \!&\sum\nolimits_{\ell=1}^L \frac{t_{\ell}}{\tilde{m}+\log _{2} \tilde{K}} \log _{2}^+\left(\frac{ E_{\max}\Gamma_{\ell}}{ t_{\ell} \sigma^2_\ell }\right) \\
{\rm{s.t.}}\;\;\;\;
& |h_{\ell,k}^e|^2 \geq \Gamma_{\ell},\ \forall \ell,k, \label{e_c_P11_h2}\\
& \eqref{e_c_P1_t1}, \eqref{e_c_P1_t2}, \eqref{e_c_P1_v}.
\end{align}
\end{subequations}
Note that the objective function is jointly concave with $t_{\ell}$ and $\Gamma_\ell$, thus the difficulty of problem \eqref{P11} consists of the non-convex constraints \eqref{e_c_P11_h2} and \eqref{e_c_P1_v}.

Let $\overline{\bm{v}}_{\ell}^{\mathrm{H}} = [\bm{v}_{\ell}^{\mathrm{H}} \ 1]$ and $\bm{q}_k = [\mathrm{diag}(\bm{g}_{\ell}^{\mathrm{H}}) \mathbf{h}^r_{\ell,k} \ {h}^d_{\ell,k}]^{\mathrm{T}}$. Exploiting the matrix lifting technique, i.e., define $\mathbf{V}_\ell= \overline{\bm{v}}_{\ell}\overline{\bm{v}}_{\ell}^{\mathrm{H}}, \mathbf{Q}_{\ell,k} = \bm{q}_{\ell,k} \bm{q}_{\ell,k}^{\mathrm{H}}$, we have $|h_{\ell,k}^e|^2 = \mathrm{Tr}(\mathbf{V}_\ell \mathbf{Q}_{\ell,k})$. 
%Thus, problem \eqref{P11} can be converted into a convex optimization problem that can be directly solved by the existing optimization solvers such as CVX \cite{cvx} by relaxing the rank-one constraint. If the optimal solution to the relaxed problem is rank-one, the optimal solution to the original problem can be recovered by matrix decomposition. On the other hand, if the optimal solution  fails to be rank-one, additional steps such as Gaussian randomization technique need to be harnessed to construct a suboptimal solution for the original problem. Here, instead, 
Besides, we convert the rank-one constraint of $\{\mathbf{V}_\ell\}$ to a difference-of-convex form and add it to objective function as
\begin{subequations} \label{P12}
\begin{align}
\!\!\!\!\! \mathop {\max }\limits_{\scriptstyle \{t_{\ell}\}, \{\mathbf{V}_{\ell}\}, \atop \scriptstyle 
\{\Gamma_{\ell}\}}  \;\; \!\! \!&\!\! \sum_{\ell=1}^L \!\!  \frac{t_{\ell} \log _{2}^+\left(\frac{ E_{\max}\Gamma_{\ell}}{ t_{\ell} \sigma^2_\ell }\right)}{\tilde{m}+\log _{2} \tilde{K}} \! - \! \frac{1}{2\rho} \! \! \sum_{\ell=1}^L (\mathrm{Tr}(\mathbf{V}_{\ell}) \!\! - \!\|\mathbf{V}_{\ell}\|_2)  \!\!\! \\
{\rm{s.t.}}\;\;\;\;
& \mathrm{Tr}(\mathbf{V}_\ell \mathbf{Q}_{\ell,k}) \geq \Gamma_{\ell},\ \forall \ell,k, \\
& \mathbf{V}_\ell \succeq 0,\ \forall \ell, \\
& \eqref{e_c_P1_t1}, \eqref{e_c_P1_t2},
\end{align}
\end{subequations}
where $\rho > 0$ is the penalty factor, and $\|\mathbf{V}_{\ell}\|_2$ denotes the spectral norm of $\mathbf{V}_{\ell}$. The factor $\rho$ is harnessed to penalize the violation of the rank-one constraint. Note that $1/2\rho \rightarrow \infty $ when $\rho \rightarrow 0$, hence the obtained solution satisfies the rank-one constraint. 
%Note that it is not effective to initialize $\rho$ to be a very small value since the objective value will be dominated by the penalty term and  thus the term related to computation rate will be diminished in this case. Instead, initializing $\rho$ to be a sufficiently large value is a practically desirable way to obtain a good starting point for the proposed algorithm. 
%By gradually decreasing the value of $\rho$, a solution that satisfies the rank-one constraint can be obtained within a predefined accuracy. 
For any given $\rho$ and any given point $\mathbf{V}_{\ell}^{(i)}$ in the $i$-th iteration, by linearizing $\mathrm{Tr}(\mathbf{V}_{\ell}) \! - \!\|\mathbf{V}_{\ell}\|_2$ to $\mathrm{Tr}((\mathbf{I} - \bm{\lambda}(\mathbf{V}_{\ell}^{(i)})  \bm{\lambda}^{\mathrm{H}}(\mathbf{V}_{\ell}^{(i)}))\mathbf{V}_{\ell} )$ via the successive convex approximation (SCA) technique,\footnote{$\bm{\lambda}^{\mathrm{H}}(\cdot)$ is the eigenvector corresponding to the biggest eigenvalue.} problem \eqref{P12} is converted to a convex optimization problem. Hence, by decreasing the value of $\rho$ and updating $\mathbf{V}_{\ell}^{(i)}$ in each iteration, the convergence of proposed algorithm for (P1) can be finally reached \cite{MengHuaDynamic}.

\subsection{Proposed Algorithms for (P2)}

For (P2), we reveal the following proposition to simplify the original problem and then propose an efficient algorithm.

\begin{proposition}\label{proposition2}
The optimal solution of (P2) satisfies
	\begin{align}
	\min_k (p_{\ell,j,k}^*|{h}^e_{\ell,j,k}|^2) = e_{\ell,j}/t_{\ell,j}^* \min_k |{h}^e_{\ell,j,k}|^2,\ \forall \ell,
	\end{align}
where $e_{\ell,j}$ denotes the maximum energy of the devices in cluster $\ell$ allocated for IRS BF pattern $j$, which satisfies $\sum\nolimits_{j=1}^J e_{\ell,j} \leq E_{\max}, \ \forall \ell$. 
%\begin{subequations}
%\begin{align}
%	&e_{\ell,j} = t_{\ell,j} \max_k p_{\ell,j,k},\ \forall \ell,j, \\
%	&\sum\nolimits_{j=1}^J e_{\ell,j} \leq E_{\max}, \ \forall \ell.
%\end{align}
%\end{subequations}
\end{proposition}

\quad {\it{Proof:}}  Suppose that $\mathcal{S}^*= \big\{ \{t_{\ell,j}^*\}, \{\bm{v}_{j}^*\}, \{p_{\ell,j,k}^*\}\big\}$ achieves the optimal solution to (P2), we have $p_{\ell,j,k}^* t_{\ell,j}^*\leq e_{\ell,j},\ \forall \ell,j,k$. The maximum allowed transmit power for the devices in cluster $\ell$ under BF pattern $\bm{v}_{j}^*$ becomes $P_0 = e_{\ell,j}/t_{\ell,j}^*,\  \forall \ell$. By combing it with \eqref{e_MSE_opt1}, Proposition \ref{proposition2} is obtained.~$\hfill\blacksquare$

According to Proposition 2, (P2) can be reformulated as 
\begin{subequations}
\begin{align}
\mathop  {\max }\limits_{\scriptstyle \{t_{\ell,j}\}, \{\bm{v}_{j}\}, \atop \scriptstyle 
\{e_{\ell,j}\}}  \;\; \!\! \!&\sum\nolimits_{\ell=1}^L \sum\nolimits_{j=1}^J \frac{ t_{\ell, j}\log _{2}^+\! \left( \! \frac{  e_{\ell,j} \min_k |h^e_{\ell,j,k}|^2}{ t_{\ell, j} \sigma^2_\ell } \! \right)}{\tilde{m}+\log _{2}  \tilde{K}} \\
{\rm{s.t.}}\;\;\;\;\;\;
& \sum\nolimits_{j=1}^J e_{\ell,j} \leq E_{\max},  \label{e_c_P21_e} \\
& \ \eqref{e_c_P2_t1}, \eqref{e_c_P2_t2}, \eqref{e_c_P2_v}.
\end{align}
\end{subequations}
By introducing the auxiliary variables $\{\gamma_{\ell,j}\}$ that satisfy $  \min_k |{h}^e_{\ell,j,k}|^2 \geq \gamma_{\ell,j},\ \forall \ell$, we have 
\vspace{-5pt}
\begin{subequations} \label{P14}
\begin{align}
\mathop  {\max }\limits_{\scriptstyle \{t_{\ell,j}\}, \{\gamma_{\ell,j}\}, \atop \scriptstyle 
\{e_{\ell,j}\}, \{\bm{v}_{j}\}}  \;\; \!\! \!&\sum\nolimits_{\ell=1}^L \sum\nolimits_{j=1}^J \frac{ t_{\ell, j} \log _{2}^+\left(\frac{ e_{\ell,j} \gamma_{\ell,j} }{ t_{\ell,j} \sigma^2_\ell }\right) }{\tilde{m}+\log _{2} \tilde{K}} \\
{\rm{s.t.}}\;\;\;\;\;\;
& |{h}^e_{\ell,j,k}|^2 \geq \gamma_{\ell,j},\ \forall \ell,j,k \label{e_c_h2} \\
& \eqref{e_c_P2_t1}, \eqref{e_c_P2_t2}, \eqref{e_c_P2_v}, \eqref{e_c_P21_e}.
\end{align}
\end{subequations}
Introducing slack variables $S_{\ell,j}= e_{\ell,j} \gamma_{\ell,j},\  \forall \ell,j$, it yields 
\vspace{-5pt}
\begin{subequations} \label{p2_1}
\begin{align} 
\mathop  {\max }\limits_{\scriptstyle \{t_{\ell,j}\}, \{\gamma_{\ell,j}\}, \atop \scriptstyle 
\{S_{\ell,j}\}, \{e_{\ell,j}\}, \{\bm{v}_{j}\}}  \;\; \!\! \!&\sum\nolimits_{\ell=1}^L \sum\nolimits_{j=1}^J \frac{ t_{\ell, j} \log _{2}^+\left(\frac{ S_{\ell,j} }{ t_{\ell,j} \sigma^2_\ell }\right) }{\tilde{m}+\log _{2} \tilde{K}} \\
{\rm{s.t.}}\;\;\;\;\;\;
& S_{\ell,j} \leq e_{\ell,j} \gamma_{\ell,j},\  \forall \ell,j, \label{e_c_S_lj} \\
& \eqref{e_c_P2_t1}, \eqref{e_c_P2_t2}, \eqref{e_c_P2_v}, \eqref{e_c_P21_e},  \eqref{e_c_h2}.
\end{align}
\end{subequations}
Since the objective value can always be increased by raising $S_{\ell,j}$ until the constraint  \eqref{e_c_S_lj} becomes active, it can be seen that the constraint  \eqref{e_c_S_lj} is satisfied with equality for the optimal solution to problem \eqref{p2_1}.

Observing that the objective function of problem \eqref{p2_1} is show as a concave function with respect to the variables $S_{\ell,j}$ and $t_{\ell, j}$, but the constraints \eqref{e_c_P2_v}, \eqref{e_c_h2}, and  \eqref{e_c_S_lj} are still non-convex. First, we convert \eqref{e_c_S_lj} into
\vspace{-2pt}
\begin{align} \label{eq16}
	S_{\ell,j} \leq \big( (e_{\ell,j} + \gamma_{\ell,j})^2 - (e_{\ell,j}^2 + \gamma_{\ell,j}^2) \big)/2,\  \forall \ell,j,
\end{align}
and then apply SCA technique to linearize the right-hand sides of \eqref{eq16}  with given points $\{e_{\ell,j}^{(i)}, \gamma_{\ell,j}^{(i)}\}$ in the $i$-th iteration as
\vspace{-2pt}
%\begin{align} 
%	\! \!  \! \! \big( \!  ( e_{\ell,j}^{(i)} \!+\!  \gamma_{\ell,j}^{(i)})^2 \!\! +\!\!   (\!  e_{\ell,j}^{(i)} \! \!+\!\!   \gamma_{\ell,j}^{(i)})(\!  e_{\ell,j} \! \! +\! \gamma_{\ell,j} \!  - \!\!   e_{\ell,j}^{(i)} \! - \! \gamma_{\ell,j}^{(i)}\! ) \! - \!(\! e_{\ell,j}^2 \! + \! \! \gamma_{\ell,j}^2\! )\!  \big)/2. \! \! 
%\end{align}
\begin{align} 
	 &\big( ( e_{\ell,j}^{(i)} +  \gamma_{\ell,j}^{(i)})^2  +   2( e_{\ell,j}^{(i)}  +   \gamma_{\ell,j}^{(i)})( e_{\ell,j}  + \gamma_{\ell,j}  -  e_{\ell,j}^{(i)}  - \gamma_{\ell,j}^{(i)})  \nonumber \\
	 &\quad - ( e_{\ell,j}^2 +  \gamma_{\ell,j}^2)  \big)/2. 
\end{align}
Second, we convert \eqref{e_c_h2} to
\begin{align}
	|\overline{\bm{v}}_{j}^{\mathrm{H}} \bm{q}_k |^2 \geq \gamma_{\ell,j}, \ \forall \ell,j,k,
\end{align}
%\begin{align}
%	|{h}^d_{\ell,k} + \bm{v}_{j}^{\mathrm{H}} \mathrm{diag}(\bm{g}_{\ell}^{\mathrm{H}}) \mathbf{h}^r_{\ell,k}|^2 = |{h}^d_{\ell,k} + \bm{v}_{j}^{\mathrm{H}} \bm{q}_k|^2 = |\overline{\bm{v}}_{j}^{\mathrm{H}} \bm{q}_k |^2,
%\end{align}
%where $\overline{\bm{v}}_{j}^{\mathrm{H}} = [\bm{v}_{j}^{\mathrm{H}} \ 1], \bm{q}_k = [\mathrm{diag}(\bm{g}_{\ell}^{\mathrm{H}}) \mathbf{h}^r_{\ell,k} \ {h}^d_{\ell,k}]$.
and linearize it as
\begin{align}
\vspace{-2pt}
	2\Re\{ (\overline{\bm{v}}_j^{(i)})^{\mathrm{H}} \mathbf{Q}_{\ell,j,k} \overline{\bm{v}}_j \} - (\overline{\bm{v}}_j^{(i)})^{\mathrm{H}} \mathbf{Q}_{\ell,j,k} \overline{\bm{v}}_j^{(i)} \geq \gamma_{\ell,j},
	\vspace{-2pt}
\end{align}
in which $\overline{\bm{v}}_j^{(i)}$ is the given point in iteration $i$. Furthermore, problem \eqref{p2_1} is approximated as a convex optimization problem by loosening \eqref{e_c_P2_v} to $|\overline{v}_{j,n}| \leq 1,\ \forall j,n$ which can be addressed by the standard solvers, the corresponding unit-modulus BF patterns are then obtained by subtracting the phases.

The proposed algorithm for tackling the dynamic IRS BF scheme is shown as solving a series of convex problems. 
Specifically, problem \eqref{p2_1} has five kinds of variables, the last of which has dimension $JN$, and the others are $LJ$. Hence, the corresponding computational complexity is given by ${\cal O}\left( I_{\mathrm{iter}}J^{3.5}(4L+N)^{3.5} \right)$ via the interior-point method \cite{Interior_point_Book}, in which $I_{\mathrm{iter}}$ stands for the quantity of iterations required to attain convergence.
Note that the amount of constraints and optimization variables scale linearly with the number of participated clusters $L$ and BF patterns $J$, it can be observed that the related computational complexity is extraordinarily large. 

\begin{remark}
	By analyzing (P2), we can observe that the optimal association between IRS BF patterns $\{\mathbf{\Theta}_j\}$ and clusters is binary  \cite{DynamicWPCNs}, i.e., each cluster only needs to select one BF pattern to compute the function. 
\end{remark}
% By analying (P2), we can observed that the optimal association between the IRS BF patterns $\{\mathbf{\Theta}_j\}$ and the clusters is binary  \cite{DynamicWPCNs}, each cluster only needs to employ one BF pattern to compute the function. 

\textbf{Low-complexity algorithm for (P2)}: Inspired by Remark~1 and the cluster-adaptive BF scheme, we can propose an efficient low-complexity algorithm. We assume that each cluster is assigned with a specific BF as the cluster-adaptive scheme presented in Section III-A.The clusters are then arranged in descending order by their respective computation rates. Each of the first $J\!-1$ clusters is given with a dedicated BF pattern and the remaining $L\!-\!J+1$ clusters are assumed to employ the same BF pattern $\mathbf{\Theta}_J$. Finally, we jointly optimize the time allocation and $\mathbf{\Theta}_J$. 
By relaxing the norm-one constraints of IRS BF, the computational complexity of the first phase is given by  ${\cal O}(LN^{3.5})$ \cite{Interior_point_Book}. The maximum computation complexity of the sorting algorithm is given by ${\cal O}(L^{2})$. With given IRS BF, we have $\eta_{\ell} = E_{\max}/t_{\ell} |h^e_{\ell, \min}|^2, \; 1\leq \ell \leq J-1$ where $|h^e_{\ell, \min}|^2 = \min_k |h^e_{\ell,k}|^2$. Reformulate the problem as \eqref{P14}, the remaining variables to be optimized is $ \bm{v}_J, \{t_{\ell}\}, \{\gamma_{\ell}\}$, where the dimensions are given by $N, L, L-J+1$, respectively. By linearizing the non-convex constraint respect to $\{\gamma_{\ell}\}$ via SCA technique and relaxing the norm-one constraint, the problem can be converted to a convex one. Hence, the overall computational complexity of this low-complexity algorithm is given by ${\cal O}(LN^{3.5}+L^2+ I_{\textrm{iter}}(2L-J+N+1)^{3.5} )$ \cite{Interior_point_Book}.

\section{Simulation Results}
In this section, we provide numerical results to demonstrate the efficiency of the proposed schemes and the insights for IRS-aided multiple-cluster AirComp system. The AP and the IRS are located at $(0,0,10)$ meter (m) and $(10,0,10)$ m, respectively. 
The placement of each device is random within a radius of $10$ m that centered at $(10, 10, 0)$ m. 
The path-loss exponents of the AP-devices channels are set to 3.3, whereas the path-loss exponents for the AP-IRS and IRS-devices channels are set to 2.3. 
Additionally, the signal attenuation at a reference distance of $1$ m is set as $30$ dB. Unless otherwise stated, other parameters are set as: $K_\ell = \tilde{K} = 5, \ \forall \ell$, $L = 5$,  $E_{\max} = 0.01$ J, $\sigma_z^2= -80$ dBm, and $T_t=0.1$ s. For comparison, we consider the following schemes: 1) \textbf{Upper bound}: relaxing the rank-one constraint in problem \eqref{P12}, which serves as a performance upper bound; 2) \textbf{Cluster-adaptive BF}: the approach in Section III-A to solve (P1); 3) \textbf{Dynamic BF}: the approach in Section III-B to solve (P2); 4) \textbf{Low-complexity algorithm}: the proposed low-complexity algorithm for (P2).

{\color{blue}
\begin{figure}[ht]
\vspace{-3pt}
\centering
\includegraphics[width=0.85\columnwidth]{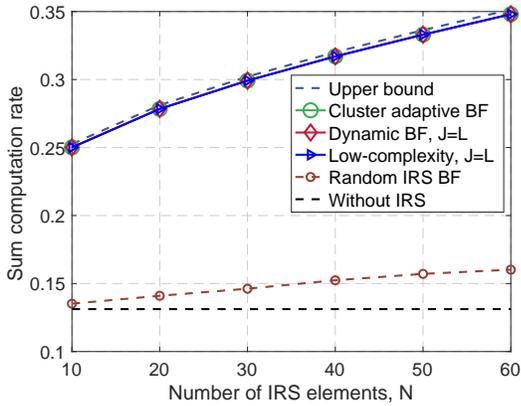}
\vspace{-5pt}
\caption{Computation rate versus the number of IRS elements $N$.}
\label{fig_simu_1}
\vspace{-5pt}
\end{figure}

\begin{figure}[ht]
\vspace{-3pt}
	\centering
	\includegraphics[width=0.87\columnwidth]{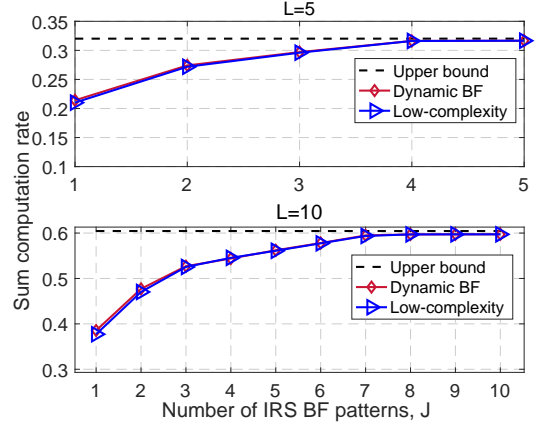}
	\vspace{-5pt}
	\caption{Computation rate versus the number of IRS BF patterns $J$.}
	\label{fig_simu_2}
	\vspace{-8pt}
\end{figure}
}

In Fig. \ref{fig_simu_1}, we analyze the system performance versus the number of IRS elements $N$. It can be seen that the dynamic IRS BF designs in the case of $L=J$ can perform almost as well as the special case with cluster-adaptive BF and nearly as the upper bound. Additionally, it can be seen that the low-complexity algorithm is capable of achieving the same performance as the joint optimization strategy. This shows the applicability of the proposed DIBF design. In comparison to the cases that BFs are randomly given and without IRS, it is observed that the proposed algorithms can greatly enhance computation rate, with the performance gap widening as the number of IRS elements $N$ rises. Besides, the computation rate of IRS-assisted AirComp with optimized IRS phase monotonically increases with respect to the number of IRS elements $N$ since more elements can reflect more captured signal energy.
%Besides, the IRS-aided AirComp with random IRS phase shifts outperforms that without IRS when $N$ becomes large, since the IRS is able to reflect some of the dissipated signals back to the AP. 

As shown in Fig. \ref{fig_simu_2}, we depict the computation rate versus the number of available BF patterns $J$. It can be seen that the low-complexity algorithm for DIBF can achieve similar performance as the previously proposed joint optimization algorithm,  making it more desirable in practice. In addition, one can see that the dynamic IRS BF scheme's performance improvement progressively reaches saturation as $J$ rises. Employing a total of $J = 7$ ($J = 4)$ BF patterns is virtually able to attain the maximum performance for $L=10$ ($L = 5$), and adding $J$ further only results in a marginal performance enhancement. It is further noted that the proposed DIBF design can achieve nearly optimal computation rate with a negligible performance loss (less than $6\%$) with only half the number of the maximum BF patterns, demonstrating its potential to meet the performance-cost tradeoff by managing the number of BF patterns.

\section{Conclusion}
In this paper, we studied the IRS-aided multi-cluster AirComp system and proposed two types of IRS BF schemes, which struck a balance between the system performance and cost. By jointly optimizing the time allocation for various clusters, the phase shifts at IRS, and the power allocation at devices, the sum computation rate maximization problems for the two scenarios were addressed. 
We proposed both general and low-complexity algorithms to tackle the optimization problem with any number of IRS BF patterns, and thus provided considerable flexibility in balancing between the performance gain of DIBF and its consequent system cost. 
Numerical results demonstrated that IRS can significantly improve the function computation performance, and presented the performance-cost tradeoff for IRS-aided AirComp. In particular, the dynamic IRS BF scheme was validated to be a cost-efficient strategy to achieve close-to-optimal performance with only half number of IRS BF patterns, and the proposed low-complexity algorithm can achieve satisfactory performance compared to the generic design.

\bibliographystyle{IEEEtran}
\bibliography{refs} 

\balance

\end{document}